\documentstyle{espart}
\input epsf
\begin{document}
\begin{frontmatter}
\title{A Water Tank \v{C}erenkov Detector for Very High Energy Astroparticles} 
\author{P. Bauleo},
\author{A. Etchegoyen\thanksref{CONI}},
\author{J.O. Fern\'andez Niello\thanksref{CONI}},
\author{A.M.J. Ferrero},
\author{A. Filevich\thanksref{CONI}},
\author{C.K. Gu\'erard\thanksref{SC}},
\author{F. Hasenbalg},
\author{M.A. Mostaf\'a},
\author{D. Ravignani},
\author{and J. Rodr\'{\i}guez Martino}.

\address{Departamento de F\'{\i}sica, Comisi\'on Nacional de
Energ\'{\i}a At\'omica, Avenida~del~Libertador~8250, (1429) Buenos
Aires, Argentina}

\thanks[CONI]{Fellow of the Consejo Nacional de Investigaciones
Cient\'\i ficas y T\'ecnicas,  Argentina}
\thanks[SC]{Permanent address: Department of Physics, University of
South Carolina, SC 29208, USA}

\begin{abstract}

Extensive airshower detection is an important issue in current
astrophysics endeavours.  Surface arrays detectors are a common
practice since they are easy to handle and have a 100\% duty cycle.  In
this work we present an experimental study of the parameters relevant
to the design of a water \v{C}erenkov detector for high energy
airshowers.  This detector is conceived as part of the surface array of
the Pierre Auger Project, which is expected to be sensitive to ultra
high energy cosmic rays.  In this paper we focus our attention in the
geometry of the tank and its inner liner material, discussing pulse
shapes and charge collections.

\end{abstract}       

\end{frontmatter}

\newpage

\section{\bf Introduction}

\v{C}erenkov light occurs when a charged particle passes through a
transparent dielectric material with a velocity greater than the
velocity of light in that material, thus producing a cone of
electromagnetic radiation whose emission  angle is related to the
velocity of the particle. Particle detectors based on this effect are
used in the examination of fast particles emitted from the atmosphere
due to the passage of a high energy cosmic ray primary, giving rise to
an extensive airshower. Such airshower might produce in turn
\v{C}erenkov light when traversing an appropriate medium.  A further
use is in particle physics mainly concerning with particle velocity or
particle mass identification in experiments involving the scattering of
light relativistic particles.

The detection of cosmic rays, which are known to reach the earth from
all directions, has occupied scientists since their discovery at the
begining of this century. Of particular interest are the very
high-energy cosmic rays\cite{bird} (in the range of 10$^{15}$ to
10$^{20}$ eV)\cite{eas}. Apart from present experiments, there are some
projects, like the Pierre Auger Project\cite{dr} which focuses its
interest in the very upper limit of this range of energies, having in
mind the identification of the primaries and sources. The project aims
at building two observatories, one in each hemisphere, over an area of
3 000 km$^2$ each, with atmospheric fluorescence and surface
detectors.

A commonly used surface detector for such particles is a water
tank\cite{hp}, where the \v{C}erenkov light produced by the impinging
shower particles is registered by an array of photomultiplier tubes.
The water tanks are simple, easy to maintain, require little
electronics, and have a large sensitivity to showers at large zenith
angles.
         
The purpose of such a tank is to measure the signal left by particles
and separate (at least statistically), the muons from the
electromagnetic contents of the shower based on the fact that the
amount of \v{C}erenkov light produced by a muon is different in average
from that produced by either an electron or gamma ray, due to their
difference in average kinetic energy (see \cite[page 111]{dr}, and
bremsstrahlung. The tanks should produce similar signals for equivalent
incoming particles, irrespective of their entrance angle and position
on the lid, therefore giving as a constraint the need for a highly
diffusing inner liner material for the tank.

This work presents the design, construction and operation of a water
\v{C}erenkov tank,  the first built to scale 1:1, as a design model for
the Pierre Auger Project. There are other prototypes of this sort
currently in operation elsewhere\cite{agasa,fnal}. In section 2 we
describe the main features of the design which ends up in a versatile
instrument. In section 3, the results from the different experiments
carried out in order to characterize the detector are presented.  A
discussion on the performance of the detector is given in section 4.

\section {\bf Mechanical Design and Experimental Setup}

A schematic diagram of the detector is shown in Fig. 1.  It is a
cylindrical water tank 1.85 m tall and 10 m$^2$ surface area.  The
material of the wall, floor and lid  is polished, non-magnetic 304
stainless steel with a thickness of 0.68 mm. All weldings were made in
continuous electric  fused seams, with no addition of solder material.
The temperature was carefully controlled during welding to avoid
modification of the chemical properties of the steel, for instance,
resistance to rusting which would impair the water absorption length.
Indeed, no signs of corrosion are observed either inside or outside
the tank after 18 months of operation.  The tank has a simple flat lid
which can be easily removed and reinstalled by a single operator with
the help of a small hand-driven hydraulic crane. After some trials, a
good darkening condition was achieved by a double layer of black cloth
ribbon installed around the border of the lid. The inner band is glued
to the lid, the outer one is kept in place by means of two elastic
cords. A conical plastic cover was installed over the tank to prevent
rainwater accumulation on the lid, which may fall into the tank when
opening. Panel feedthrough connectors were installed on the tank wall,
about 10 cm over maximum water level in order to feed the HV to the
photomultiplier tubes (PMT) and receive the detector signals.
 
The light emitted as \v{C}erenkov radiation is detected by
photomultiplier tubes: on reaching the photocathode, a \v{C}erenkov
photon might produce the ejection of an electron, thus called
photoelectron. The maximum possible water depth is 1.6 m, which allows
to completely enclose 8" Hamamatsu 1408 PMT's inside the tank. A
flexible system composed of two fixed and two rotary aluminum holder
bars permits positioning the tubes at different angles and at any
radii, and also to independently vary their height. These features
enable to use the instrument to check the results of simulations, for
instance assuming different geometrical configurations for the
detector. Three PMT's have been mounted vertically in the tank, with
their photocathodes immersed in the water surface, looking down. Their
operating voltages were 1260 V for PMT's 1 and 3, and 1300 V for PMT 2.
This geometry was chosen because simulations (see \cite[page 155]{dr})
indicate that positioning the PMT's at the bottom, looking up, would
produce a highly anisotropic collection of light: particles entering
the tank towards a PMT would generate \v{C}erenkov photons, and those
generated close to exiting the tank concentrate towards the tube.
Although the number of photons per unit track is essentially constant,
those produced on entrance will be spread over a large area on reaching
the bottom, whereas those at the particle exit would reach the bottom
close to the particle direction.

The water tank has been placed on a platform that allows positioning of
plastic-scintillator counters under the tank, at any radius. Also
counters were placed on the tank, so as to trigger events with
coincidence signals between upper and lower counters and thus the muon
trajectories are defined.  Counter sizes are 23 $\times$ 40 cm$^2$ (top
counters), and 15 $\times$ 15 cm$^2$ (bottom counters).

The inner surface of the tank was degreased, washed with water and mild
detergent and rinsed abundantly with the same water quality to be used
as detector material, obtained from a reverse osmosis water treatment
plant. Water resistivity at the outlet of the plant was 250
k$\Omega$~cm and after the filling hose it was reduced to about 200
k$\Omega$~cm.  The purpose of using reverse osmosis water instead of
resin-interchanged deionized water was to explore the adequacy of this
type of water for its use in Auger's \v{C}erenkov detectors in which
preservation of the water transparency is required for a very long
time. Obviously, the former type of treatment implies economic
advantages over the latter.

A recirculation system, consisting of a 0.75 HP plastic body
centrifugal pump and a 10 $\mu$m Micronite filter, was installed by the
detector, connected by 3/4" black plastic hoses to diametrically
opposite points in the tank wall, 10 cm above the bottom. The
recirculation system is operated, in average, only one day per week and
very little amount of material is usually retained in the filter. No
additives of any type were added to the water.

An inner tank diffusive liner was placed in order to improve light
collection uniformity for different incident muons. For such a purpose
we installed a 0.1 mm thick Tyvek\cite{Tyvek} material on both inner
walls and tank floor. Details of tests of this liner will be shown in
section 3.2. The Tyvek was supported by 3 aluminum rings pressing
against the tank wall.  At the tank bottom it was stretched over a
circular aluminum structure which hangs from six nylon strings.  In
this way, by changing the vertical position of this Tyvek floor it is
easy to simulate any desired effective water depth in the tank.
Finally, a removable Tyvek disk with holes in the desired PMT's
positions, was installed at the tank top by simply floating it on the
water surface.

Pulses coming out from the PMT's anodes were carried to the acquisition
room (15 m away from the tank), using RG213 low attenuation cables. We
currently use a CAMAC charge-sensitive ADC (LeCroy 2249A), having a
sensitivity of 0.25 pC/channel to record the output of each PMT. We
have also acquired PMT wavefronts on a digital oscilloscope (Tektronix
THS 720) of 500~MSPS.

\section {\bf Measurements and Experimental Results}

The current detector measures  \v{C}erenkov photons with wave lengths
ranging from 300 to 600 nm, due to the PMT quantum efficiency
bandwidth. Photons might undergo three kinds of interactions in the
tank, mainly with the electrons of the medium\cite{pp}: i) elastic
Rayleigh scattering, ii) absorption, or iii) medium boundary
interactions. Elastic scattering, although not negligible, is less
important than absorption.  For instance, at $\lambda = 350$ nm the
absorption length is 21 m while the scattering length is 97 m for the
clearest waters \cite{smi81}. Moreover, after being scattered the light
would still be present in the tank, whereas it disappears after
absorption.

\subsection {\sl PMT's Calibration}

In order to characterize and optimize the response of this \v{C}erenkov
detector a series of measurements has been performed. Since our key
interest is to study both pulse shape and number of photoelectrons, a
calibration of the phototubes was required. This calibration is done
measuring the  single photoelectron spectrum of the PMT's, that is, the
spectrum of the charge collection at the PMT anode by events in which a
single photoelectron was emitted by the cathode.

Therefore, to ensure single-electron detection, the photocathodes were
covered with aluminium foils with a $\approx$ 1 cm$^2$ hole since it
was experimentally observed that under this condition 90 $\%$ of the
events showed no signal output from the phototubes. The remaining 10
$\%$ of the events are accompanied by one or more photoelectrons, being
of course the probability of a single photoelectron the largest one.
This was experimentally checked during the course of this calibration.

The mean anode charges from a single photoelectron for each of the
three PMT's were 3.2, 2.5 and 2.2~pC, with an estimated error of 1
channel (0.25 pC). With these values, the number of photoelectrons in
an actual experiment with the PMT's fully uncovered (ie. without
aluminium foils) is just the ratio of the collected charge to the
latter numbers. An example of the response of the PMT's in a real
experiment is shown in Fig. 2 where the mean values of the collected
charge were 145, 64, and 97~pC. These charges correspond to a mean
number of photoelectrons of 45, 26, and 44.

A thorough check of the present calibration would be to reproduce the
shape of the fully uncovered-PMT spectra by means of the obtained
single photoelectron spectra and of simulated \v{C}erenkov photons. For
this purpose we developed the following method of combining
experimental and simulated results. We simulated the \v{C}erenkov
photons produced by the passage of vertical central muons which fully
traverse the tank.  The \v{C}erenkov photons were created by Montecarlo
calculations with the computer code GEANT \cite{geant}. Since both the
water absorption length and angular distribution of the reflected light
from the tank surface are not experimentally known, these photons were
not tracked through the tank. It seems more appropriate to assume that
the number of photoelectrons emitted at the PMT's follows a Poisson
distribution for any given number of \v{C}erenkov photons produced,
with its mean value given by the experiment.

From the generated \v{C}erenkov photons distribution, we randomly
choose a number of photons, and
then, from the Poisson distribution we again randomly obtain the number
of photoelectrons.  Finally, using the experimental data of the anode
charge produced by this given number of photoelectrons, a theoretical
histogram of simulated charge distribution was generated (see solid
line in Fig. 2). In short, this is a three step process: i) a given
number of generated photons is chosen at random from the GEANT
spectrum, ii) a Poisson distribution is asummed centered at the
experimental value of produced photoelectrons multiplied by the ratio
of the number of photons chosen in i) and the mean value of the GEANT
distribution, and then a given number of generated photoelectrons is
randomly chosen from this distribution, iii) each of these
photoelectrons undergo a random sampling of the single photoelectron
spectra. It is worth mentioning that the number of simulated muons is
the same as the number of experimentally detected events, thus no free
paramenters enter in this calculation. The mean value of this
calculation is expected to coincide with data due to the way the
Poisson distributions were produced and as such there is an overall
normalization.  Still, it is emphasized the good shape agreement to the
data which justifies the three processes assumed.

A second method to evaluate the number of photoelectrons has also been
used. It is based on the fact that the charge spectrum width is related
to the number of  photoelectrons in the PMT. This stems from the three
random processes mentioned above since the observed width can be
estimated by adding in quadrature the width of these three
distributions\cite{dra}, which after some algebra yields: 

\[
 \left( {\sigma\over mean} \right)^2_{exp} \; = \;
 {1.5\over<PE>}\, +\, 0.093^2 
\]
where $<PE>$ is the mean number of photoelectrons.

This method has the quite important advantage over the previous one
that is independent of the experimental setup like PMT gain, although
its uncertainties (20\%) are larger than in the more conventional
approach (10\%). Both methods predict consistent results. It is also
noted that this consistency even applies for the PMT which has the
reduced number of photoelectrons.

\subsection {\sl Characterization Studies}

Several studies were carried out in order to characterize the detector:
a comparison between stainless steel and Tyvek liner, an optimization
of the position of the PMT's at the detector top and of the tank depth,
and a comparison between black and Tyvek tops.

One key detector design issue is whether there is a need for a liner
material, covering the inner part of the tank. Therefore, we performed
two sets of relative measurements using a blue--LED light source, whose
emission was limited by an opaque shade to a 45$^\circ$ cone: firstly
with the bare polished stainless steel surfaces and secondly with the
Tyvek liner.  The LED was pointing downwards, placed at the water
surface, 1.5 m above the bottom of the tank. The results are shown in
Fig. 3, where the collected light intensity is plotted as a function of
the radial PMT's position.  The peak appearing in the stainless steel
data is interpreted as a geometrical effect of the specular reflection
of the light, which concentrates there due to the cylindrical shape of
the wall. This effect is similar to the caustic curve observed in
optics and can easily be accounted for by simulations as shown in ref.
\cite{exp}. As can be seen in the figure the use of the Tyvek liner
dramatically uniformizes light collection.  Curves a) and b) have
different normalization since the aim of this experiment was to study
solely the uniformity of light.  Nevertheless, the parameter of
interest for the normalization is the percent reflectance which ranges
from 38 to 54 \% and from 70 to 90 \% for stainless steel and Tyvek,
respectively for the bandwidth 300-600 nm\cite{irv}.

In this work we are also interested in the optimization of the radial
position of the PMT's on the upper lid and the tank depth. Regarding
the former, the main purpose was to find a PMT arrangement such that
the \v{C}erenkov light collection would be as independent as possible
of the particles entrance position. The obvious symmetry of the tank
prescribes that the three PMT's should be 120$^\circ$ apart. We
selected 7 vertical muon entrance positions by making use of the
scintillator counters (see Fig. 4): one at the central point, three at
$r = 80$ cm and three at $r = 160$ cm, at 0$^\circ$, 30$^\circ$, and
60$^\circ$, from a PMT. The PMT's were setup at radii 80, 120, and 160
cm, therefore, allowing the collection of data for 21 experimental
setups. The effective tank depth was 1.20 m. The results of these
measurements are summarized in Table I.  Regarding the position of the
PMT's two issues are to be taken into account: the total charge
collection and its uniformity, ie. the mean value and its dispersion.
It can be seen that the greater charge collection is obtained with the
PMT's at 80~cm, and that the charge collection uniformity does not
significantly vary within the experimental errors for any fixed PMT
radial position.

Apart from optimizing the PMT's position on the horizontal plane, this
versatile prototype permits to vary the effective depth of the water by
raising the Tyvek floor. There is a compromise between tank depth and
track length. Indeed deeper tanks would augment the number of
\v{C}erenkov photons for vertical traversing particles but on the other
side would increase the average track length (ie. more absorption) for
reaching a PMT. Therefore the optimum depth has to be experimentally
found. Different measurements have been performed with water depths,
ranging from 30 to 150 cm, with muons vertically entering through the
tank centre (see Fig. 5).  It is apparent from the figure that there is
a continuos rise in the number of photoelectrons as a function of the
water depth. A comparison should also include electromagnetic particles
which are mostly stopped in the tank in actual experimental conditions.
This can not readily be done with an experimental setup consisting in a
single prototype since no trigger would be available. Some simulations
were performed\cite{muem} for vertical particles and, as expected,
electrons (and gamma's after conversion) in average produce smaller
signals with tank depth due to absorption and to the fact they cannot
benefit with a longer track because of their lower mean kinetic energy
and bremsstrahlung. As such, there is a tendency to separate better
muons from electromagnetic particles.
 
We also performed a signal study with either a Tyvek or a black top. An
illustration of pulse shapes, averaged over 200 events, is presented in
Fig. 6, for a single PMT. In full line a typical mean pulse is shown:
it has a very short rise time ($\approx 10$ ns) and a longer decay
time. Such a rise short time would probably arise from \v{C}erenkov
photons reflecting from the floor or hitting the wall one extra time
before reaching the PMT. The decay time corresponds to light traversing
the tank several times. For example, since the light speed in water is
22.5 cm/ns, a photon arriving 200 ns later would have had a total track
inside the tank of 45 m. Drawn in dotted line is the mean pulse
corresponding to a black top condition, ie replacing the Tyvek floating
top by a black polyethylene foil. As expected, the pulse amplitude is
esentially the same since it should have no contribution from light
bouncing off the top. However, the decay time, and therefore the charge
collection, is greatly diminished. The decay times, extracted from an
exponential fit to the tail of the spectra (also shown in the figure),
are 11 and 39~ns, whereas mean charges are 49 and 112~pC for black and
Tyvek tops, respectively (note that the estimate of the ratio of
photoelectrons from these two numbers, 2.29, is in agreement within
errors to the ratio obtained from the width of the distributions, 2.26,
as obtained from the formulae of section 3.1). There appears to be a
compromise between shorter decay times and larger charge collection but
the reduction in charge for energetic muons is of lesser significance
since the amplitude of the pulse does not change and therefore no
signals are lost merged in the background. The charge collected,
normalized by the mean values, are quite similar as appreciated in
Fig.7. The peaks are essentially the same, just being the black top one
slightly wider, due to the reduced number of photoelectrons collected.

\section{\bf Conclusions}

We have undertaken an experimental study of different design parameters
of a typical water \v{C}erenkov tank, as those to be used as surface
detectors of the Pierre Auger Project.
 
The most important design consideration of this model detector is its
flexibility to adapt to a variety of experiments, since it has a lid
that permits to install new equipment and to easily change the
arrangement of the PMT's in the tank and the effective detector depth.

Airshower \v{C}erenkov detectors should have little light absorption
and have similar responses for different entrance positions of
particles to the tank. In this work we have experimentally shown that
the Tyvek liner material much better complies with the uniformity
requirement than stainless steel, at least for blue-LED light. Further
to this, it was shown for vertical muons that the Tyvek response
guarantees uniformity, and therefore different placements of the PMT's
on the detector top would yield similar performances. This is not
necessarily the case for the dependence of charge collection on PMT's
radial position.  We have also shown that deeper tank depth will
produce larger electronic signals for energetic muons. It is always
better to have bigger tanks not only for the above-mentioned reason but
also for the larger solid angle coverage of the sky (a larger tank,
however, would increase costs and manufacturing difficulties).
 
It is worth mentioning, that our conclusion on  a more preferable
deeper tank stems from the large total absorption length relative to
the tank depth due to the water purity ( 20-30 m of average absorption
coefficient in our tank)  and for the large value of the Tyvek
reflection coefficient.

Another consideration is related to pulse shapes, ie decay time and
charge collection. The black top experiment gave a dramatic difference
in these two items: decay lengths of 11 and 39~ns and charge
collections of 49 and 112~pC for black and Tyvek tops, respectively.
But, since the experimental pulse amplitudes do not vary, no extra
black-top collected pulses are lost in the backgroud.

We are grateful to Dr. C. Hojvat for his contributions with the CAMAC
electronics. We also thank A. Boschan and J.I.  Franqueiro for their
collaboration during part of the measurements and J. Vidall\'e and J.
Fern\'andez V\'asquez for their help.

\vfill\eject
\centerline{\bf {Tables}}

\noindent {{\bf Table I}: Ratio of mean charge collected over the three
PMT's to the respective value for vertical central muons with PMT's
placed at 120 cm.}

\vskip 1cm

\begin{tabular}{|ll||ccc|}
\cline{1-5}
\multicolumn{2}{|c||}{Entering} &\multicolumn{3}{c|} {PMT's position}\\ \cline{3-5}
\multicolumn{2}{|c||}{Position} & 80~cm & 120~cm & 160~cm \\
\cline{1-5}
R =  0 cm  & $\theta = 0^\circ$  & 1.31 & 0.92 & 0.63 \\
R = 80 cm  & $\theta = 0^\circ$ & 1.27 & 0.95 & 0.72 \\
R = 80 cm  & $\theta = 30^\circ$ & 1.06 & 0.95 & 0.79 \\ 
R = 80 cm  & $\theta = 60^\circ$ & 1.35 & 1.05 &  \\
R = 160 cm & $\theta = 0^\circ$  & 1.05 & 0.99 & 0.71 \\
R = 160 cm & $\theta = 30^\circ$ & 1.20 & 1.02 & 0.60 \\
R = 160 cm & $\theta = 60^\circ$ & 1.05 & 0.86 & 0.69 \\
\cline{1-5}
\multicolumn{2}{|c||}{Mean Value} & 1.2 & 0.96 & 0.69\\
\multicolumn{2}{|c||}{Dispersion} & 0.1 & 0.06 & 0.06\\
\cline{1-5}
\end{tabular}

Note that the normalization experiment was done with a different set of
measurements yielding a ratio of 0.92 rather than 1.0.

\begin{figure}
\epsfysize=18.0truecm
\epsffile{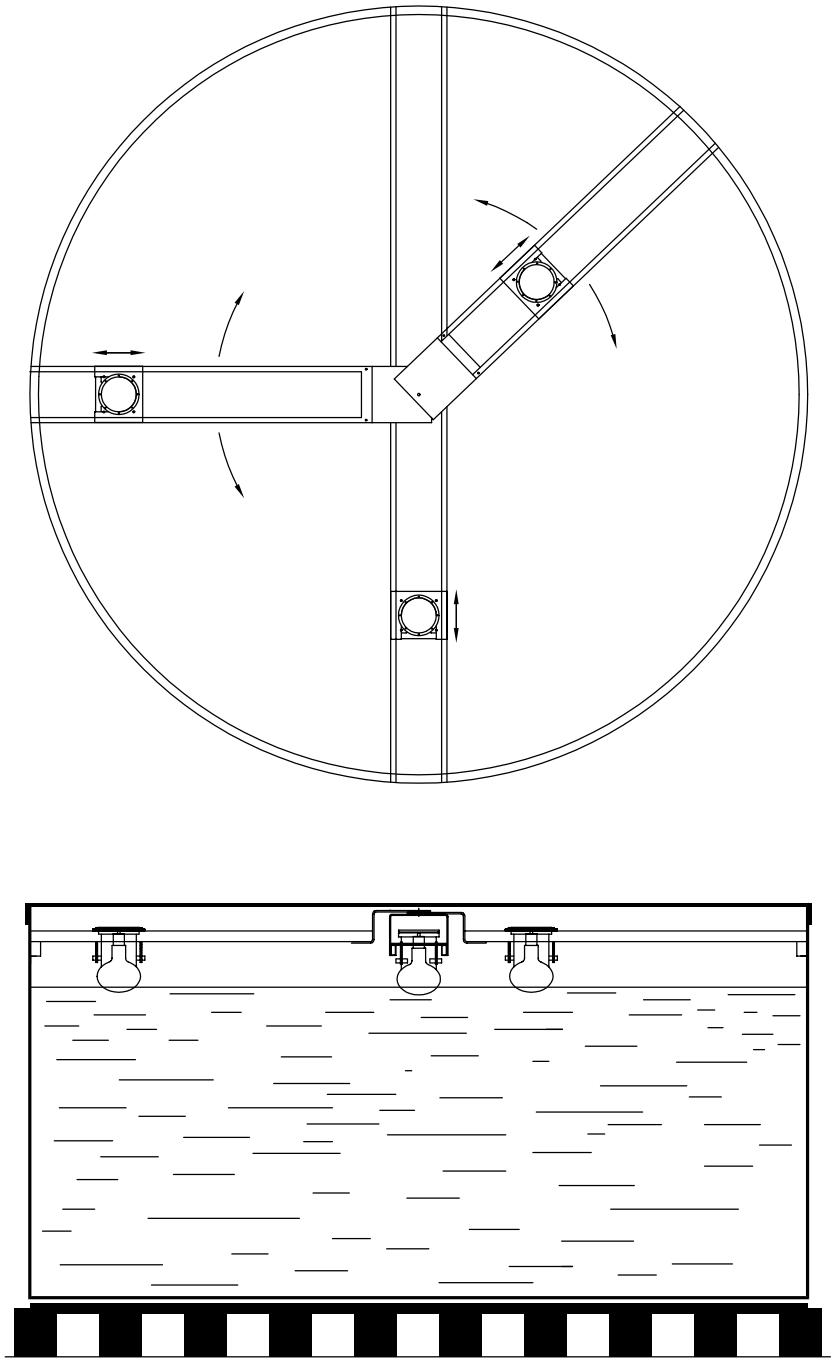}
\caption {a) Top view of the PMT's holders. The arrows indicate
possible movements of the two rotatable arms and radial displacements
of the PMT's, which can therefore be placed in any desirable
geometrical arrangement; b) Side view of the tank. Actual dimensions
are 10 m$^2$ $\times$ 1.85 and the maximum attainable water depth is
1.6 m.}
\end{figure}

\begin{figure}
\epsfysize=20.0truecm
\epsffile{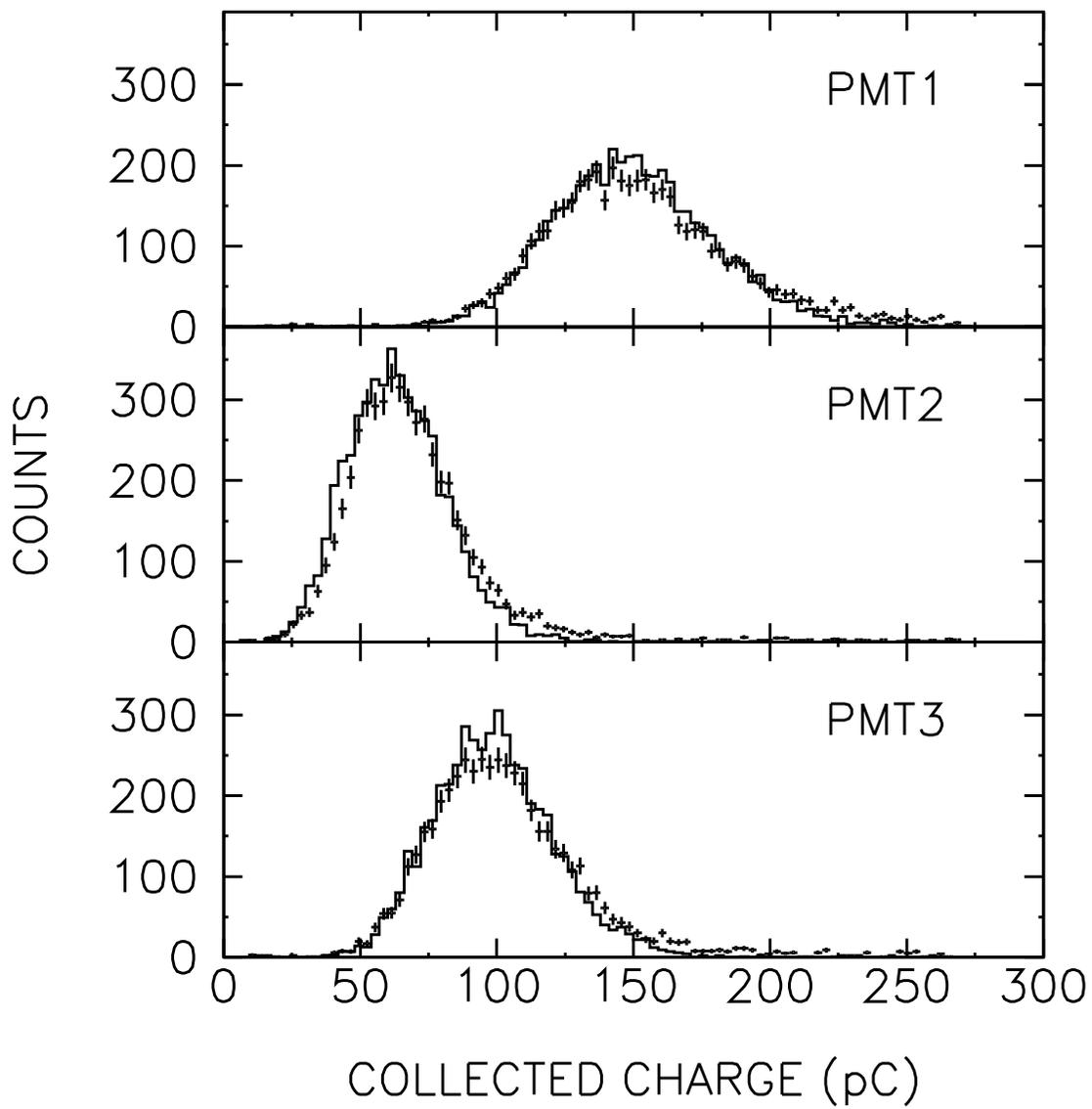}
\caption {Mean charge collected for central muons for each PMT. The
dotted line corresponds to a simulated peak using the obtained single
photoelectron charge collection.}
\end{figure}

\begin{figure}
\epsfysize=20.0truecm
\epsffile{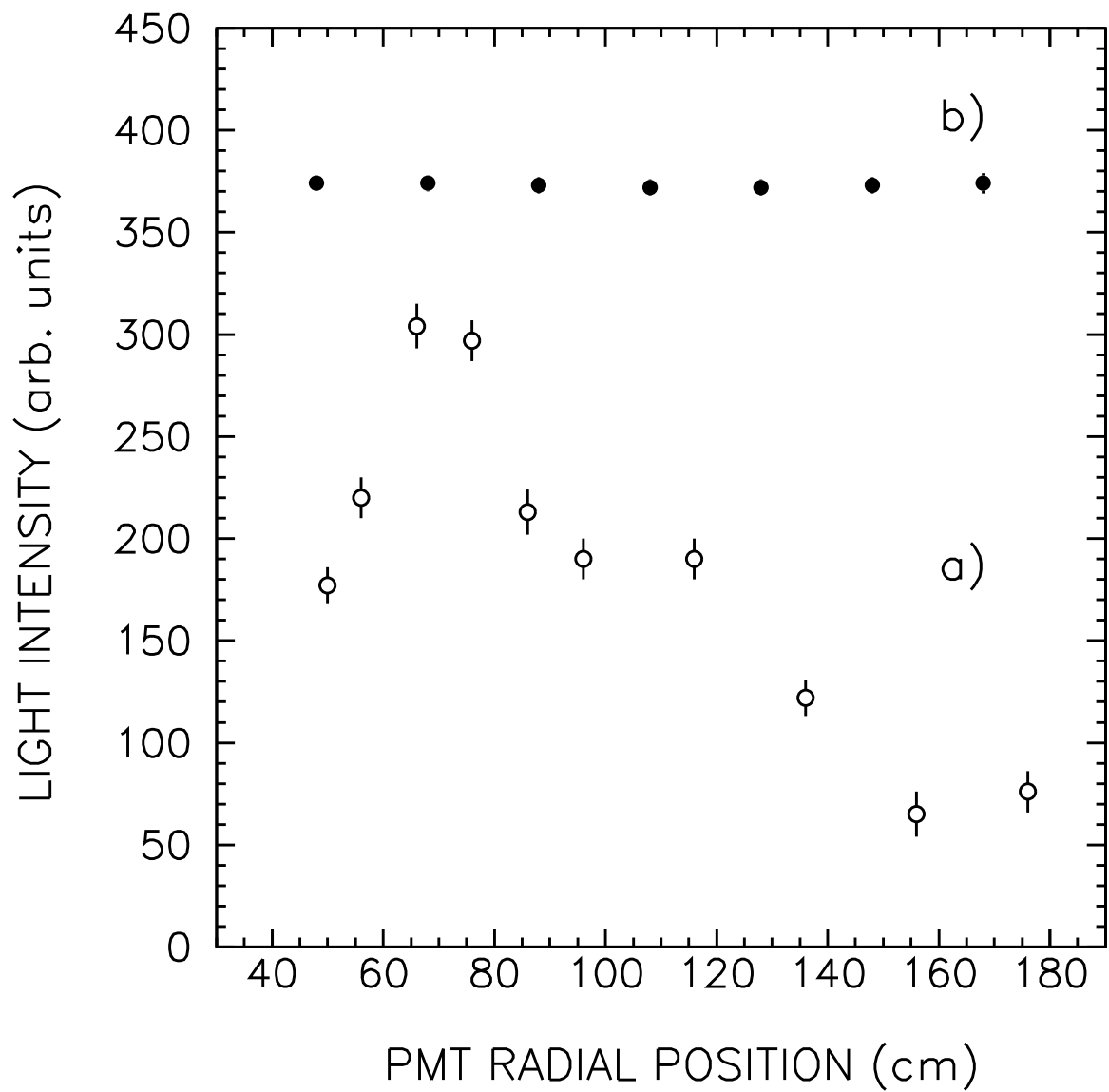}
\caption {Charge collected in a PMT due to a LED--light source vs.
radial PMT position; a) with stainless steel walls, b) with Tyvek
liner. The arbitrary units are different for plots a) and b) since the
experimental normalizations are different.}
\end{figure}

\begin{figure}
\epsfsize=19.cm
\epsffile[-40 0 450 760]{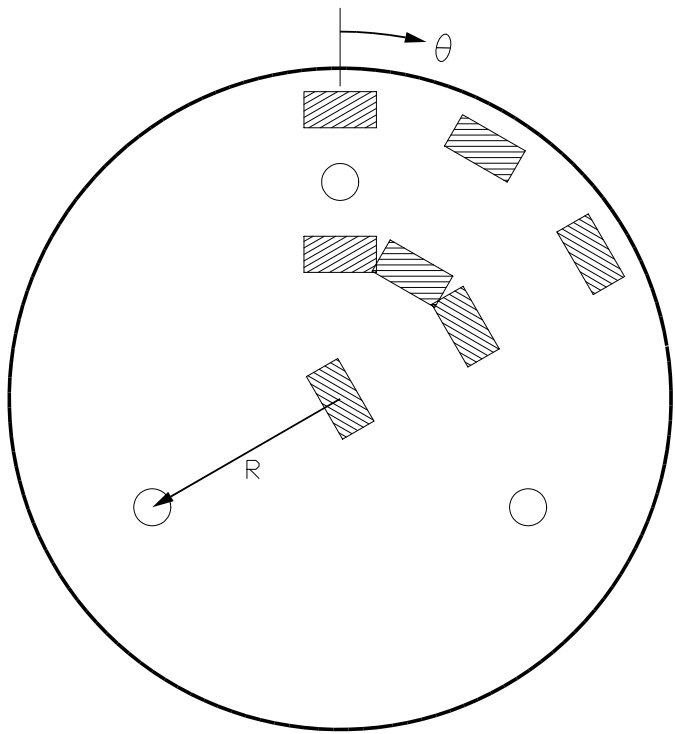}
\caption {Setup for charge-collection amount and uniformity
measurements for vertical muons.  In dashed rectangles are shown the 7
counter positions, both upper and lower, and in open circles one of the
PMT locations. The drawing is to scale with PMT's and upper counters.}
\end{figure}

\begin{figure}
\epsfysize=20.0truecm
\epsffile{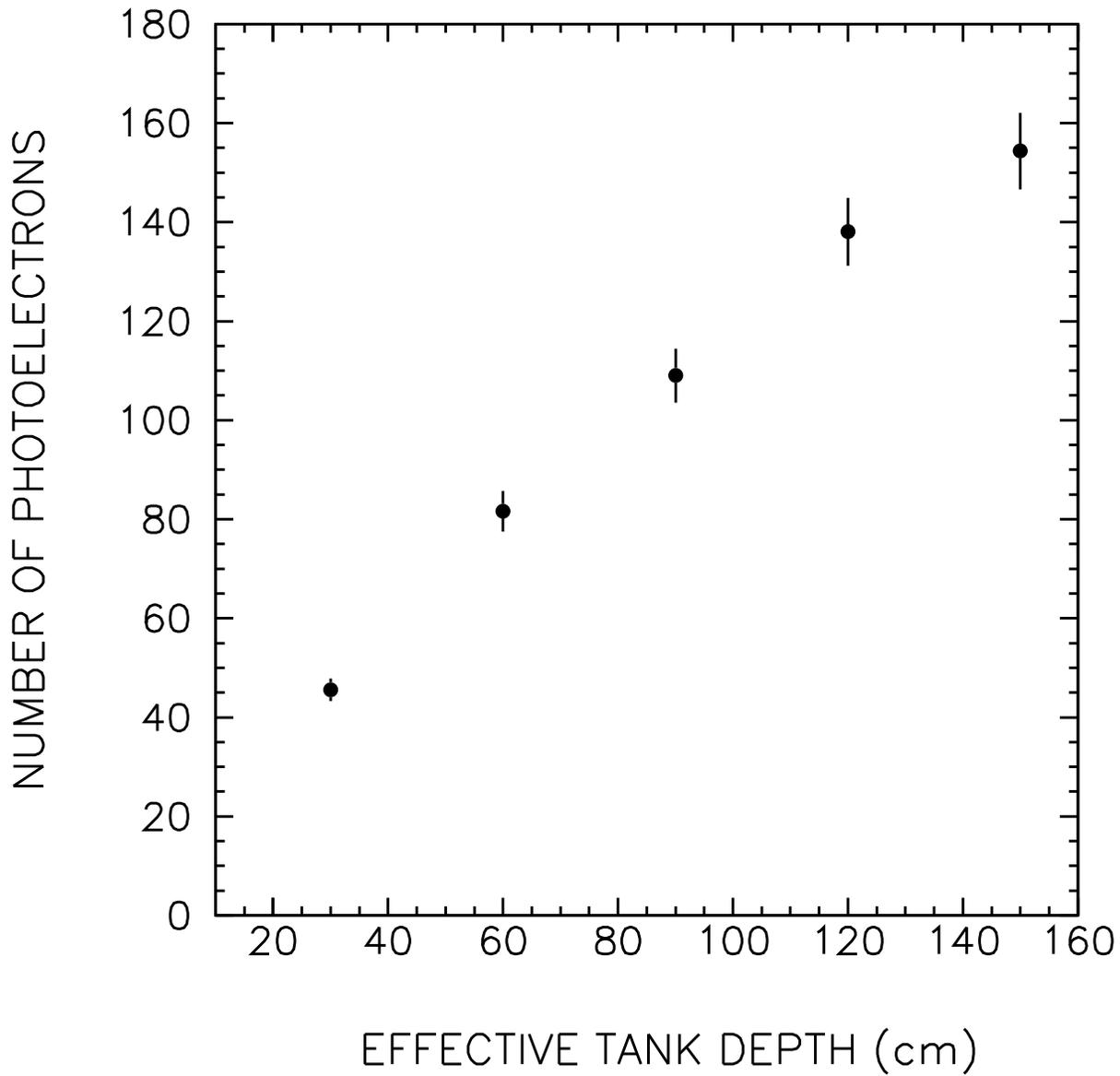}
\caption {Number of photoelectrons as a function of the effective tank
depth.  The photoelectrons correspond to the sum over the three PMT's,
which were located at R = 120 cm.}
\end{figure}

\begin{figure}
\epsfysize=20.0truecm
\epsffile{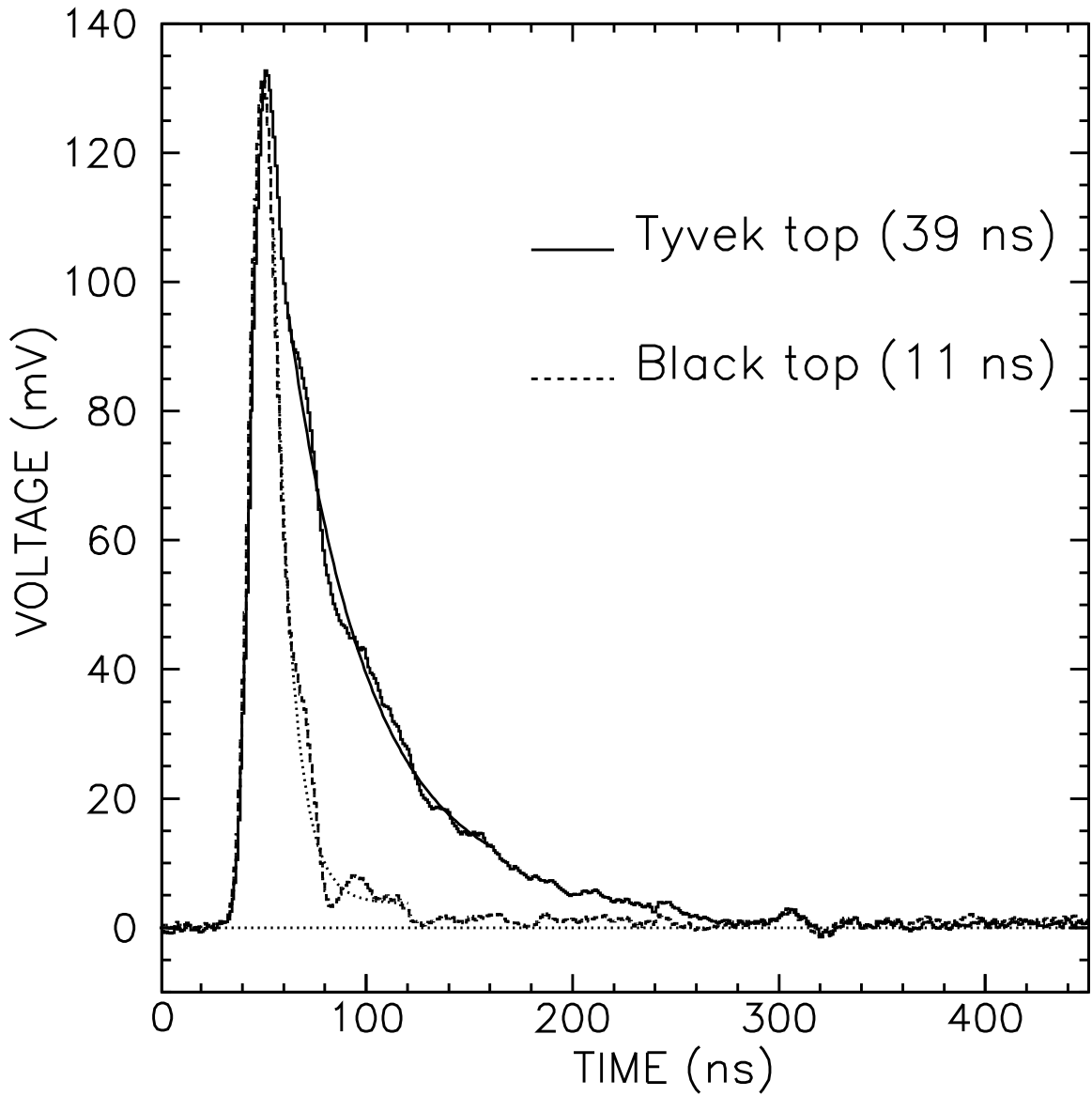}
\caption {Average pulse shapes for a single PMT. The full line
histogram corresponds to a fully covered inner tank with Tyvek liner
whereas the dotted line histogram to a black liner top. Also shown are
exponential fits to the decay times.}
\end{figure}

\begin{figure}
\epsfysize=20.0truecm
\epsffile{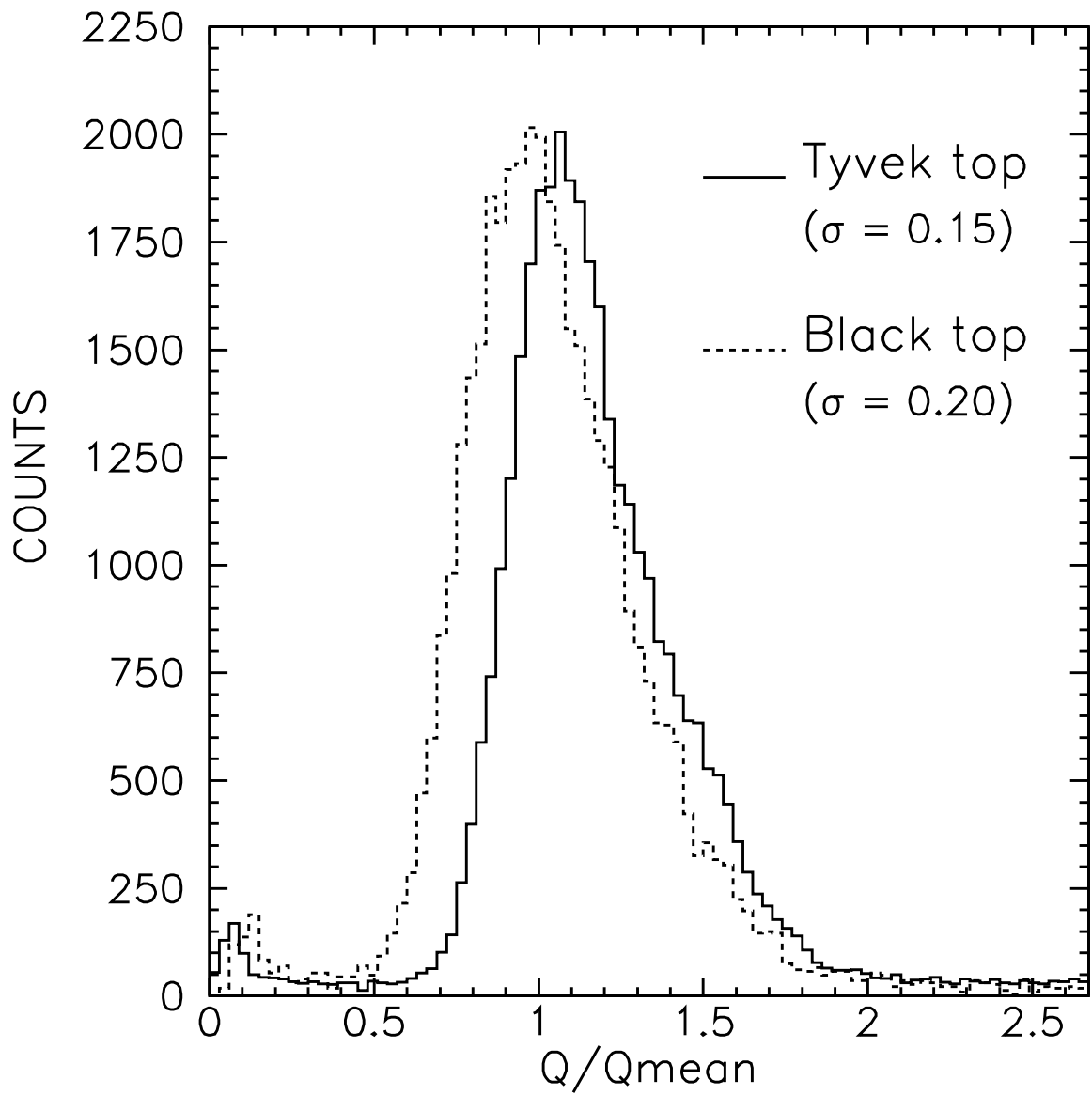}
\caption {Total charge collected for a single PMT, normalized to its
mean value.  Full and dotted line histograms correspond to Tyvek and
black tops, respectively. The sigma values were obtained from a
gaussian fit precluding the right-hand-side non-gaussian tail.}
\end{figure}

\end{document}